\documentclass[aps,amsmath,prl,twocolumn,a4paper,showpacs,floatfix]{revtex4}
\usepackage{graphicx}
\newcommand{\comment}[1]{}
\newcommand{\beq}{\begin{equation}}
\newcommand{\eneq}{\end{equation}}
\newcommand{\bea}{\begin{eqnarray}}
\newcommand{\enea}{\end{eqnarray}}
\newcommand{\bc}{\begin{center}}
\newcommand{\ec}{\end{center}}

\newcommand{\met}{\frac{1}{2}}

\newcommand{\h}{\hbar}
\newcommand{\bd}{\begin{displaymath}}
\newcommand{\ed}{\end{displaymath}}

\newcommand{\bwt}{\begin{widetext}}
\newcommand{\ewt}{\end{widetext}}
\begin{document}

\title{  Quantum interference  of  electrons  in  a 
ring: tuning of the geometrical phase} \author{R. Capozza$^{1,*}$}
\author{D. Giuliano$^{1,2}$} \author{P. Lucignano$^{1,3,4}$} \author{A. Tagliacozzo$^{1,3}$}

\affiliation{$^1$ Dipartimento di Scienze Fisiche Universit\`a degli
             studi di Napoli "Federico II ", Napoli, Italy}
\affiliation{$^2$ Dipartimento di Fisica, Universit\`a della Calabria and
             I.N.F.N., Gruppo collegato di Cosenza, Arcavacata di Rende
             I-87036, Cosenza, Italy}
\affiliation{$^3$ Coherentia-INFM,
Monte S.Angelo - via Cintia, I-80126 Napoli, Italy }
\affiliation{$^4$  SISSA and 
             INFM Democritos National Simulation Center, Via Beirut
             2-4, 34014 Trieste, Italy}

\date{\today}
\begin{abstract}
We calculate the oscillations of the DC conductance across a
mesoscopic ring, simultaneously tuned by applied magnetic and electric
fields orthogonal to the ring. The oscillations depend on the
Aharonov-Bohm flux and of the spin-orbit coupling. They result from
mixing of the dynamical phase, including the Zeeman spin splitting, and of 
geometric phases. By changing the applied fields, the geometric 
phase contribution to the conductance oscillations can be tuned
 from the adiabatic (Berry) to the nonadiabatic (Ahronov-Anandan) 
regime. To model a realistic device, we also include nonzero backscattering 
at the connection between ring and contacts, and a random phase for electron 
wavefunction, accounting for dephasing effects.

\end{abstract}
\maketitle

In mesoscopic quantum devices, the wavefunctions of charged particles
may acquire a nonzero phase, when undergoing a closed path in a space
threaded by external fields. For instance, electrons traveling in an
external magnetic flux $\phi $ pick up an Aharonov-Bohm (AB) phase
\cite{aharonovbohm}, which can be read out from DC conductance
oscillations in an interference device \cite{webb}.  Also, spin-orbit
interaction (SOI) couples orbital and spin electronic degrees of
freedom, thus giving rise to an effective, momentum dependent, field,
which adds a geometric (topological) \cite{berry,anandan} phase to the
electron wavefunction
\cite{gefen1,loss,dario,aronov}.

Recently, it has been shown that SOI can be controlled by means of
voltage gates in III-V semiconducting mesoscopic structures (Rashba
effect) \cite{meijer,miller,noi}. This has aroused a renewed interest
in studying transport in ballistic rings, in the presence of Rashba
coupling \cite{morpurgo,nitta,frustaglia,molnar,souma}. Yet, it is
still controversial under which conditions the spin dynamics
adiabatically follows the orbital motion in a device like this and
whether the Berry phase can be detected in the oscillations of the
transmission altogether \cite{yau}.  Also, it is, up to now, still
unclear, what are the possible consequences of dephasing due to small
fluctuations of the length of the arms, or scattering at the
connections between the device and the leads.

\begin{figure}[!htp]
    \centering
    \includegraphics[width=\linewidth]{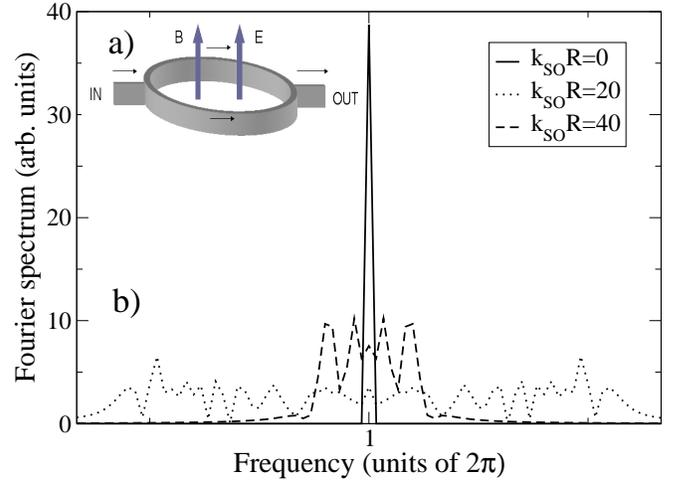}
\caption{(color on line) 
         a) Sketch of the device studied, 
         b)Fourier transform of the conductance {\sl vs} of Fig.(\ref{esp}
         [right panel]) 
           for $k_{\text {SO}} R=0, 20,40$. The variable
         conjugate to the magnetic flux $\phi/\phi _0 $ is  in units of
         $2\pi $} 
\label{ftr}
\end{figure}
In this paper, we report extensive results concerning ballistic
quantum transport across a $1d$ ring, in the presence of both an
orthogonal magnetic field and of SOI.  We compute the DC conductance
by means of the Landauer formula \cite{landauer} $ G={e^2}/{\h}
\sum_{\sigma\sigma'} \left|A(\sigma;\! \sigma'|E)\right|^2$, 
where $A(\sigma ;\sigma'|E)$ is the probability amplitude for an
electron entering the ring with energy $E$ and spin polarization
$\sigma '$ to exit with spin polarization $\sigma $.  We employ a
real-time path integral approach \cite{feynman}, and we use the saddle
point approximation for the orbital motion (which singles out an
optimum constant velocity for the electron, $\dot\varphi $).  At each
contact, the trasmission is weighted with an amplitude $\bar{t}\;
e^{iz}$, and the reflection takes place with amplitude is $\bar{r}\;
e^{iz}$, where $z$ is a stochastic variable with flat distribution in
$[-\zeta , \zeta ]$, which encodes dephasing effects. Eventually, we
average $N$ times over different dephasing realizations.  The winding
in the ring before escaping provides the electron propagator with an
extra phase, which includes the combined effect of ``geometrical'' and
``dynamical''phases, arising from AB, SOI and Zeeman spin splitting
(ZSS) (proportional to the cyclotron frequency $\omega_c$).  Our
approach applies to any regime, either adiabatic, or nonadiabatic, as
the spin propagator is evaluated exactly. In the limiting regimes, in
which ZSS is either much larger, or much less, than SOI, the dynamical
and the geometrical phases can be easily identified.  The intriguing
regime is the non adiabatic one, when ZSS and SOI are of comparable
strength.  

\begin{figure}[!htp]
    \centering
    \includegraphics[width=\linewidth]{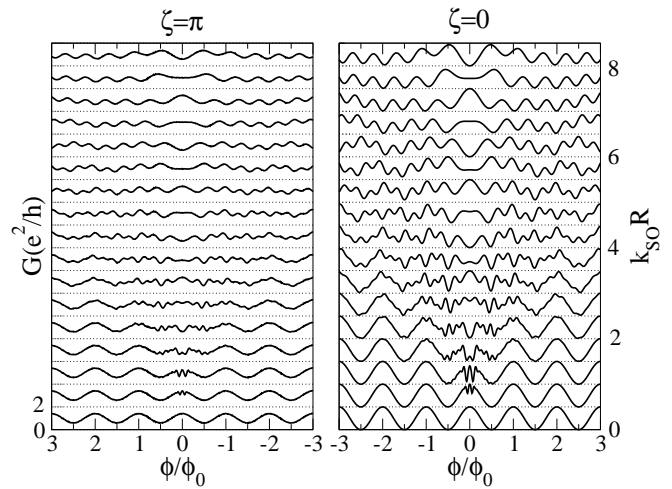}
\caption{[Right panel] Conductance vs $\phi / \phi_0$ for increasing values of
SOI at $\zeta = 0$  and $\bar{r} =0.05$. 
[Left panel] The same plot as at the right panel, but 
for  $\zeta = \pi$.  }
\label{esp}
\end{figure}

In Fig.(\ref{ftr} b), we plot the Fourier transform of the
interference contribution to the DC conductance for three increasing
values of SOI (Fig.\ref{esp}[right panel]), with very little back
reflection at the connections between ring and leads, and no dephasing
($\zeta =0 $).  In the absence of SOI (solid line), we see only the
peak corresponding to AB oscillations. At increasing SOI strength
(dotted line), more structures appear, which eventually evolve into a
four-peak structure for a larger value of SOI (dashed line). The
four-peak feature confirms the interpretation by Yau {\sl et al.}
\cite{yau} and supports the conclusion that the Berry phase can be
detected experimentally in similar devices.  We consider the dynamics
of a spinful single electron injected at the Fermi energy in a ring
with equal arms \cite{sigrist}, as sketched in Fig.(\ref{ftr} a). In
calculating the transmission in an orthogonal electric and magnetic
field, we neglect the actual finite transverse dimension of the arms
of the ring, as this would alter the result only quantitatively
\cite{moroz}.   Our model Hamiltonian is given by:

\beq
H = 
\frac{\hbar^2}{2mR^2}\left(\hat{{l}}+\frac{\phi}{\phi_{0}}\right)^2
+\met \hbar
\omega_{c}\:\sigma_{Z}+
\frac{\alpha}{\hbar}\left( {\hat z} {\times} 
\left({\vec p+\frac{e}{c}\vec A}\right) \right){\cdot}
{\vec{\mathbf \sigma}} \:
\label{hamilt}\:,
\eneq
where $ \hbar {{\hat l}}=i \hbar \partial_\varphi$ is the angular
momentum operator, $\varphi$ is the orbital coordinate along the ring,
and $ \vec{ \sigma}$ are Pauli matrices; $\alpha$ is a coupling
constant, including the effect of the electric field (in units of $eV$
\AA), $k_{\text {SO}}R = 4 \alpha \tau _0 /(\hbar R )$, where $\tau _0
= mR^2/2 \hbar $ is the time scale of orbital fluctuations (note that
$\omega_c = \tau_0^{(-1)} \phi / \phi_0$).
Since we are interested in the transmission amplitude in time $t_f$ ,
$ A ( \sigma_f , t_f ; \sigma_0 , 0)$, we sum over paths within
homotopy classes, corresponding to the electron winding $n + 1/2$
times in the ring ($n+1/2$ is positive or negative, depending on
whether the electron path winds clockwise, or counterclockwise)
\cite{moran}.  We assume ballistic quantum propagation at energy $E_0$
(referred to the Fermi energy of the contacts), which requires
integrating over all final times $t_f >0$.  Accordingly, the
transmission amplitude for an electron entering the ring at $\varphi (
0 ) $ with spin polarization $\sigma_0$ and exiting at $\varphi ( 0 )
+ \pi$, with spin polarization $\sigma_f$, for a given realization of
the random phases is given by:
\bwt
\beq
A(\sigma_f ; \sigma_0 | E_0 )  = |\bar{t}|^2 \sum_{n = - \infty}^\infty
\int_{0}^{\infty} \: d t_f \; |\bar{r}|^{2 (|n| -1 )} e^{i \sum_j^{2|n|} z_j} e^{i \frac{E_{0}
t_f}{\hbar}} \int_{\varphi ( 0 )}^{
\varphi ( 0 ) +  \pi ( 2n-1 )}  {\cal D} [ \varphi ]
\langle \sigma_f,t_f| e^{ i \int_{t_0}^{t_f} d t \: {\mathcal L} [ \varphi , 
\dot{\varphi} , t ] } |  \sigma_0, 0 \rangle \:.
\label{ampclas1}
\eneq
The $n^{th}$ partial amplitude in Eq.(\ref{ampclas1}) corresponds to
summing over paths $\varphi ( t )$ satisfying the boundary conditions
$\varphi ( t_f ) - \varphi ( 0 ) = \pi ( 2n-1 )$.  We take the
transparency at the contacts to be such that backscattered
trajectories which retrace back part of the path can be
neglected. This suppresses weak localization corrections\cite{nota0}
and Altshuler-Aronov-Spivak oscillations\cite{AAS}.
The Lagrangian in Eq.(\ref{ampclas1}) is given by:
\beq
{\mathcal{L}}[\varphi(t), \dot \varphi(t),\vec \sigma]=
\frac{m}{2}R^2\dot\varphi^2(t)-\frac{\phi}{\phi_{0}} \hbar \dot\varphi(t)
+  \frac{\alpha^2\:m}{2 \hbar^2}+ \frac{\hbar^2}{8 m R^2}-
\left[\met\hbar\omega_{c}\sigma_{z}
+ \frac{\alpha R  m \dot\varphi(t)}{\hbar}
\left( e^{-i\varphi(t)}\sigma_{+}   +e^{i\varphi(t)}\sigma_{-}\right)
\right] \:.
\label{lag}
\eneq
\ewt 
We now perform the saddle point approximation on the orbital motion.
Since in Eq.(\ref{ampclas1}) the spin is still a quantum operator, we
derive the equation of motion for $\varphi$ within the coherent state
representation for spin variables (Haldane's mapping) \cite{coher}.

\beq
\frac{d}{d t} \frac{ \partial {\cal L}}{\partial \dot \varphi(t)}  -
\frac{ \partial {\cal L}}{\partial \varphi} = 0 \Rightarrow
m R^2 \ddot\varphi(t)  = 0\label{moto1}
\eneq

Thus, the dynamics of the orbital coordinate $\varphi$ decouples from
the spin dynamics, within saddle point approximation.  The solution of
Eq.(\ref{moto1}) satisfying the appropriate boundary conditions and
parametrized by the integer $n$ is

\beq
\varphi_n ( t ) = 
\varphi ( 0 )  + {\rm sign} ( n ) \pi ( 2 | n | - 1 ) 
\left( \frac{ t }{t_f}   \right) \label{mimmo1}
\eneq
\noindent
The ultimate formula for the transmission amplitude across the ring is
given by \cite{nota}
\bwt
\bea
 A(\sigma_f;\sigma_0 | E_0)= \sqrt{ \frac{m }{ 2 \tilde{E_0} } } \; 
|\bar{t}|^2\sum_{n\neq 0,n=-\infty}
^{\infty}|\bar{r}|^{2(|n|-1)}\; e^{\sum _j^{2|n|} z_j} \: 
 e^{i\frac{mR^2}{2\hbar \:t_n}(\pi(2|n|-1))^2}
e^{-i\frac{\phi}{\phi_{0}}(\pi(2|n|-1)) {\rm sign} (n)}
e^{i E_{0}t_n /\h}\times \label{final1} \\\nonumber \times
e^{i\left[ 1 + \left (k_{\text {SO}}R \right )^2 \right ] \:t_{n}/ 16 \tau _0} 
\langle\sigma_{f}|\hat{U}_{cl}(t_n, 0)|\sigma_{0}\rangle \:  ,
\enea
\ewt
with $\tilde{E}_0 = E_0 + \hbar \left[ 1 + \left (k_{\text {SO}}R \right )^2
 \right ] / 16 \tau _0 $. 
In Eq.(\ref{final1}), the  spin  dynamics  is governed  by  the  effective  
Hamiltonian $\hat H_{\rm spin} (t)=\vec b ( t )  \cdot \vec \sigma$. 
$H_{\rm spin} ( t )$  is parametrized  by  the  angular velocity of  the  
electron rounding $n+1/2$ times in the ring, $\dot{\varphi}_n$, which is
a constant, according to Eq.(\ref{mimmo1}). $H_{\rm spin} ( t )$ is the 
Hamiltonian of a quantum spin, moving in  an effective time dependent
external magnetic field $\vec b (t)=(b_z,b_-,b_+)= (\frac{\h
\omega_c}{2}, k_{\text {SO}}R \: \hbar  \dot \varphi_{n}e^{i \varphi _n
(t)}/2, k_{\text {SO}}R \:\hbar \dot \varphi_{n}e^{-i \varphi _n
(t)}/2)$. Eq.(\ref{final1}) contains  the matrix
elements  of the  spin evolution  operator $U_{cl} (t_f , 0 )=
\hat{T} \exp [ -i \int_0^{t_f} H_{\rm spin}  ( t ) \: d t ]$, 
($\hat{T}$ is the usual time-ordering operator), between states with given 
spin polarization. Such a matrix element adds a geometrical phase to the 
total amplitude. This phase   reduces to the usual Berry phase
in the adiabatic limit \cite{berry}.

To obtain Eq.(\ref{final1}) from Eq.(\ref{ampclas1}) we have used the
steepest descent approximation.  Within the $n^{\rm th}$ topological
sector, we find that the phase of the integrand is stationary at the
time $t_n= \pi [ (2|n|-1)\tau_0 ] \sqrt{\hbar
\tau _0 /\tilde{E}_0 } $. Thus, we evaluate  the  contribution  of each term  
to  the  sum of Eq.(\ref{ampclas1}) at $t=t_n$. 
Inserting Eq.(\ref{final1}) in the  Landauer  formula  allows  us  to  
compute the linear  conductance across the ring. 

\begin{figure}[!htp]
    \centering \includegraphics[width=\linewidth]{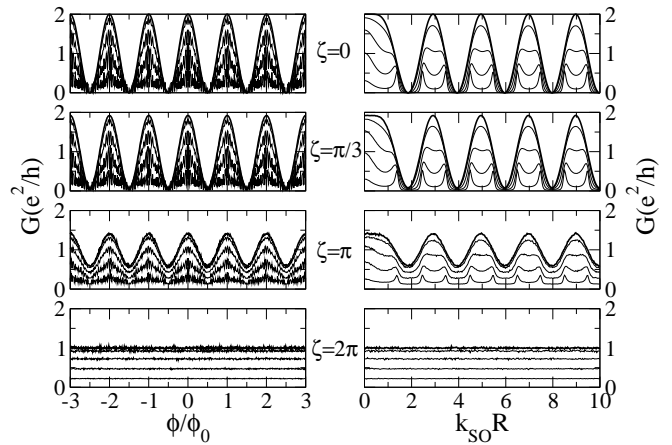}
\caption{[Left panel] Conductance  {\it vs.} 
 $\phi/\phi_0$ at $k_{\text {SO}} R=0$, for $\bar{r} =0,0.2,0.4,0.6,0.8$ 
(different  curves  from  top  to  bottom  in  each  box) and
at increasing dephasing (parametrized by $\zeta$). [Right panel]
  Conductance  {\it vs.} $k_{\text {SO}} R$ at $\phi/\phi_0= 0 $, for
the same values of $\bar{r}$ and $\zeta$.}
\label{GvsBSO}
\end{figure}

In the right panel of Fig.(\ref{GvsBSO}), we plot the DC conductance
{\it vs.}  $k_{\text {SO}}R$ at $\phi/\phi_0 =0 $ for different values
of $\bar{r} $ (different plots within a single box), and at increasing
phase randomness (boxes from top to bottom with $\zeta = 0,\pi/3 ,\pi,
2\pi$), averaged over $N=1000$ realizations. In the left panel, we
plot the DC conductance {\it vs.}  $ \phi / \phi_0$, for the same
values of $\bar{r}$ and $\zeta$, at $k_{\text {SO}}R = 0$.  In the
right panel, we see that in the case of ideal coupling, $\bar{r}=0$,
the quasiperiodic oscillation of the conductance reproduces the
localization conditions at the expected values of $k_{\text {SO}}R$
\cite{frustaglia,molnar,souma}. For $\bar{r}>0$, interference
involving winding numbers $|n + 1/ 2| >1$ gives rise to more
complicated patterns: the average and the peak value of the
conductance decrease, when the transparency of the barriers is
lowered.  The transmission is progressively reduced, when $\bar r$
increases.  Contributions from higher harmonics, due to multiple
reflections, only appear in the AB oscillations, with maximum
amplitude when $\phi / \phi_0$ is close to an integer, that is, when
the constructive interference condition is fulfilled.

We see that in both panels in Fig.(\ref{GvsBSO}) the amplitude of the
oscillations due to quantum interference are overall reduced by the
same size, because of increasing $\zeta$. Eventually they are
washed out for $\zeta=2\pi$.


In Fig.(\ref{esp}), we show the combined effect of $B$ and SOI, on the
conductance as a function of $\phi/\phi_0$ at $E_{0}=0$ and $\bar{r} =
0.05$, at increasing values of $k_{\text {SO}}R$ ( from bottom to top)
for $\zeta = 0 $[right panel], and $\zeta = \pi $ [left panel]. From
the right panel, we see that the zero-flux value of the conductance
oscillates with increasing $k_{\text {SO}}R$.  Maxima and minima are
reduced by the dephasing, as it appears from the left panel, since both
constructive, as well as disruptive interference, are suppressed. The
results at the right panel are in excellent qualitative agreement with
recent experiments \cite{nitta}. Therefore, we infer that, in real
samples the coupling between the contacts and the leads is
approximately ideal ($\bar{r} \sim 0$) and the transport is
quasi-ballistic.

The geometrical phase should be detectable as a modulation of the
interference term in the total DC conductance across the ring, on top
of the fundamental modulation due to AB-effect.  Fig.(\ref{ftr} b)
shows the Fourier transform of the patterns at the right panel of
Fig.(\ref{esp}) for $k_{\text {SO}} R = 0 , 20 , 40$. To get an
insight concerning the appearance of the four-peak feature at
$k_{\text {SO}} R = 40$, we may resort to the adiabatic approximation
for the conductance ($k_{\text {SO}}R \dot{\varphi} \ll \omega _c$),
obtaining:

\begin{eqnarray}
\sum_{\sigma \sigma'} | A ( \sigma ; \sigma' )  |^2 \approx 2 
- 2 \sum_{ \pm } \biggl\{  \cos^2 \theta  \cos \left[ 
2 \pi \frac{ \phi}{\phi_0} \pm   \pi \cos  \theta  \right]\nonumber\\
+ \sin^2  \theta  \cos  \left[ \pi \frac{ \phi}{\phi_0}
 \pm  \frac{\pi  \omega_c}{ \dot{ \varphi} } \right] \biggr\} \;\; , \;
\label{adiab}
\end{eqnarray}
\noindent
where  $\cos \theta  = \left [ 1 + \left ( { k_{\text {SO}}R 
\dot{\varphi}} /{\omega_c}  \right )^2 \right ]^{-1/2}$. 

In the absence of SOI ($\theta=0$), the former term reduces to the
usual AB-oscillating term, while the latter one simply
disappears. When SOI is $\neq 0$, but still much smaller than ZSS,
$\theta$ weakly depends on $\phi$, so that two small satellites appear
at each side of the AB peak. For $k_{\text {SO}} R = 40$, the Berry
phase becomes proportional to $\phi$. Hence, the central AB peak
splits into two, as seen in Fig.(\ref{ftr} b). Also, since $\cos^2 (
\theta )$ decreases, while $\sin^2 ( \theta ) $ increases, the
amplitude of the outer peaks (associated to ZSS) increases, while the
amplitude of the inner peaks (associated to Berry phase) decreases.
Therefore, we infer that the splitting of the AB peak into two is, in
fact, an evidence for the existence of a topological phase
\cite{yau,waghrak}.

To conclude, we have employed a path integral real time approach to
compute the DC conductance of a ballistic mesoscopic ring in both
electrical and magnetic fields. Our approach goes beyond other recent
semiclassical calculations by allowing for nonideal couplings between
ring and leads (with nonzero reflection $\bar{r}$) and for dephasing
effects. The results satisfactorily compare with experiments. 

By varying the external fields we can explore both the adiabatic and
nonadiabatic regime: the latter appears as irregular wiggles in the
middle of Fig.(\ref{esp}) [right panel].  For large Rashba couplings and
a weak magnetic field, spin flip phenomena take place, due to the
off-diagonal component of the spin evolution matrix. We stress that
$k_{\text {SO}} R =30 - 40 $ corresponds to a SOI coupling $\alpha
\sim 200\; meV \AA $, in rings with $R \sim 1\mu m$, what can be
presently achieved experimentally.  In this regime such devices can
work as spin filters.

We acknowledge  valuable discussions with D. Bercioux, G. Campagnano 
and R. Haug.

$*$ Present address: INFM-S3 e Dipartimento di Fisica, Universit\`a 
    di Modena e Reggio Emilia, via Campi 213/A, 41100 Modena, Italy.


\begin{thebibliography}{99}
\bibitem{nota0}By   scattering  at  the  contacts, we assume  that  non  
ballistic trajectories which contribute to winding number $n$ but
reflect back by retracing some of the path can be neglected.  This
applies if we are not in the extreme mesoscopic limit (i.e.  ring
diameter much smaller than the phase coherence length ), that is both
conditions, $|n|>>1 $ and $|n|(1-|r|^2) \sim 1 $ are not
simultaneously satisfied. Indeed, by our choice of the parameters,
contributions for large $|n|$ are always very small.
\bibitem{nota} Strictly  speaking  the  transmission  amplitude of 
 Eq.(\ref{final1}) should exclude the prefactor, which is the same as
 for the free particle evolution.
\end{thebibliography}
\end{document}